\documentclass[twocolumn,english,aps,prl,floatfix,showpacs]{revtex4}
\usepackage[T1]{fontenc}
\usepackage[latin1]{inputenc}

\makeatletter

\usepackage{babel}

\usepackage{babel}
\makeatother
\begin{document}

\title{Comment on {}"$Z_2$-slave-spin theory for strongly
correlated fermions\char`\"{}}

\author{Alvaro Ferraz$^{1}$, Evgeny Kochetov$^{1,2}$}

\affiliation{$^{1}$International Institute of Physics - UFRN,
Department of Experimental and Theoretical Physics - UFRN, Natal, Brazil}
\affiliation{$^{2}$Laboratory of Theoretical Physics, Joint
Institute for Nuclear Research, 141980 Dubna, Russia}


\maketitle The authors of ref.$\,$\cite{sigrist} introduce a new
slave-spin (SS) mean-field (MF) approach to the Hubbard model of strongly
correlated electrons in terms of Ising spins and standard fermion operators,
by constructing an enlarged Hilbert space with both physical
and unphysical states. The purpose of this construction is clear:
this formulation involves a much smaller number of slave fields, compared
with more standard slave particle frameworks.
In this way the complexity of the original model is reduced to a more readily tractable system
of Ising spins in a transverse field coupled to free fermions.
In this Comment, we show that the unphysical states are not properly
excluded and despite the claimed good agreement with previous results,
the consistency of their MF theory is unjustified.

The paper \cite{sigrist} may be considered
as an extension and an improvement of the formalism developed earlier in \cite{huber}.
It can also be thought of as a minimal formulation of previous slave-spin representations. \cite{05}

The single-band Hubbard model is written in the form
\begin{equation}
H=-\sum_{ij\sigma}t_{ij}c^{\dagger}_{i\sigma}c_{j\sigma}+\frac{U}{2}\sum_i(n_i-1)^2.
\label{1}\end{equation}
The hopping amplitude is denoted by $t_{ij}$ and $U$ is the onsite repulsion. The operator
$c^{(\dagger)}_{i\sigma}$ creates an electron at site $i$ with spin $\sigma$,
and $n_i=\sum_{\sigma}c^{\dagger}_{i\sigma}c_{i\sigma}.$

The authors introduce auxiliary local pseudospin variables $\vec I_i=(I^x_i,I^y_i,I^z_i)$
with eigenstates
$I^z_i|\pm\rangle_i=\pm\frac{1}{2}|\pm\rangle_i.$
The physical electron operator is then represented as
\begin{equation}
c_{i\sigma}^{(\dagger)}=2I^x_if^{(\dagger)}_{i\sigma},
\label{2}\end{equation}
where $f_{i\sigma}^{(\dagger)}$ is an auxiliary fermion operator defined on the
space spanned by the vectors $|0\rangle_i,\, |\uparrow\rangle_i,\, |\downarrow\rangle_i, \,
|\uparrow\downarrow\rangle_i$. \cite{note}

The SS representation (\ref{2}) expands the Hilbert space. The authors identify the on-site physical
Hilbert space to be spanned by the states
\begin{equation}
{\cal H}_{phys}=\{|+\rangle|0\rangle,\, |-\rangle|\uparrow\rangle,\, |-\rangle|\downarrow\rangle,\,
|+\rangle|\uparrow\downarrow\rangle\}.
\label{3}\end{equation}
The unphysical states are then
\begin{equation}
{\cal H}_{unphys}=\{|-\rangle|0\rangle,\, |+\rangle|\uparrow\rangle,\, |+\rangle|\downarrow\rangle,\,
|-\rangle|\uparrow\downarrow\rangle\}.
\label{4}\end{equation}
To obtain a faithful
representation of the original problem, the authors claim to single out the physical subspace by the local constraint
$A_i:=I^z_i+1/2 -(n_i-1)^2=0$. This equation justifies the substitution in (1)
\begin{equation}
\frac{U}{2}\sum_i(n_i-1)^2\to \frac{U}{2}\sum_i(I_i^z+1/2),
\label{5}\end{equation}
which appears as a necessary step to
recover the transverse field Ising model representation.

The projection of the Hubbard model
\begin{equation}
H'=-\sum_{ij\sigma}t_{ij}I^x_iI^x_jf^{\dagger}_{i\sigma}f_{j\sigma}+\frac{U}{2}\sum_i(I^z_i+\frac{1}{2})
\label{6}\end{equation}
onto the physical subspace
is equivalent to the original model (\ref{1}).
Such a projection can  be achieved explicitly by imposing the requirement $Q_i:=A_i^2=
1/2+I^z_i[1-2(n_i-1)^2]=0$, for each lattice site.
The projection operator
$Q_i$ commutes with Hamiltonian (\ref{1})
and generates a $U(1)$ local gauge symmetry
which takes care of the redundancy of the decomposition (\ref{2}).
However, notice that the operator constraints $A_i=0$ and $Q_i=0$ are not equivalent to each other.\cite{equiv}

The, so far, exact SS  formulation is then treated at  MF level by fully decoupling
the SS and the auxiliary fermion operators in the form
$H'\to H^{MF} = H_f+H_I,$ where
\begin{eqnarray}
H_f&=& -\sum_{ij\sigma}g_{ij}t_{ij}f^{\dagger}_{i\sigma}f_{j\sigma},
\nonumber\\
H_{I}&=&-\sum_{ij}\chi_{ij}t_{ij}I^x_iI^x_j +\frac{U}{2}\sum_i I^z_i. \label{7}
\end{eqnarray}
Here the hopping amplitude and the Ising exchange coupling are renormalized
by the factors $g_{ij}$ and $\chi_{ij}$, respectively. These are to be determined
through the self-consistency equations
\begin{equation}
g_{ij}=4\langle I^x_iI^x_j\rangle_{H_I},\,\chi_{ij}=4\sum_{\sigma}
(\langle f^{\dagger}_{i\sigma}f_{j\sigma}\rangle_{H_f}+c.c.).
\label{8}\end{equation}

To proceed, the authors replace the constraint $A_i=0$ by its ground state expectation value,
ignoring completely the condition $Q_i=0$.
In the enlarged Hilbert space, the $Q_i$ operator
is a projection matrix, $Q_i|{\cal H}_{phys}\rangle_i=0,\, Q_i|{\cal H}_{unphys}\rangle_i=
|{\cal H}_{unphys}\rangle_i.$
Because of this, the local MF constraint $\langle Q_i\rangle=0$
still eliminates the unphysical states. In contrast, the same condition does not hold for
the constraint expectation value $\langle A_i\rangle $ that can be nullified by
the unphysical subspace as well. To see this, let us pick up a state
$$|\Phi\rangle=
c_i^{(1)}|-\rangle_i|0\rangle_i +c_i^{(2)}|-\rangle_i|\uparrow
\downarrow\rangle_i$$
$$+c_i^{(3)}|+\rangle_i|\uparrow\rangle_i
+c_i^{(4)}|
+\rangle_i|\downarrow\rangle_i\in {\cal H}_{unphys},$$
where $|c_i^{(1)}|^2+|c_i^{(2)}|^2=
|c_i^{(3)}|^2+|c_i^{(4)}|^2.$
It then follows that
$\langle\Phi|A_i|\Phi\rangle=0$.
Consequently, the condition $\langle A_i\rangle =0$ no longer discriminates between physical
and unphysical states (as opposed to the requirement $\langle Q_i\rangle=0$),
and the resulting SS MF theory becomes inconsistent. The substitution (5)
now implies that the uncontrolled unphysical degrees of freedom contribute to the theory as well.

The authors refer to their MF approach as a $Z_2$ gauge theory.
They claim that the local $Z_2$ transformations,
$f_i\to u_if_iu_i=-f_i, \, I^x_i\to u_iI^x_iu_i=-I^x_i,\, u_i=1-2Q_i,$ respect the MF decomposition (\ref{7})
and "in this sense, the mean-field ansatz breaks the $U(1)$ gauge theory down to $Z_2$".\cite{sigrist}
This statement may cause some confusion. \cite{u}
It is of course legitimate to restrict the space of variational states to a subspace generated by the states related
by the $u_i$ transformations. However, this does not result
in a $Z_2$ invariant MF theory: the gauge-related
variational states still belong to the {\it different } Hamiltonians, $H^{MF}$ and $u_iH^{MF}u_i\neq H^{MF}.$
In fact, a $Z_2$ gauge invariant MF theory only arises if one considers both functions $\chi_{ij}$  and $g_{ij}$ as
Hubbard-Stratanovich dynamical fields, $\chi_{ij}=\bar\chi_{ij}\sigma_{ij},\,g_{ij}=
\bar g_{ij}\sigma_{ij}.$ Here $\bar\chi_{ij}$ and $\bar g_{ij}$ are precisely the solutions to the saddle point Eqs.(\ref{8}),
and $\sigma_{ij}=\pm 1$ is an Ising dynamical gauge field that takes care of
the $Z_2$ gauge fluctuations beyond the saddle point approximation.


\begin{thebibliography}{99}
\bibitem{note} In the limit $U\gg t$, the physical electron operators admit representations in terms of the
$su(2)$ spins and spinful fermions \cite{w}. Within that approach, the underdoped Hubbard model reduces
to the Kondo-Heisenberg problem \cite{pfk}.
\bibitem{equiv} The operator $A_i$ does not commute with the Hamiltonian (\ref{1}). The consistency
of the constraint $A_i=0$ with time evolution requires that $[A_i,H]_{{\cal H}_{phys}}=0.$
The consistency of this secondary constraint implies in turn that $Q_i=0$. Therefore the constraint
$A_i=0$ can be imposed self-consistently only if the requirement $Q_i=0$ holds.
\end{thebibliography}
\end{document}